\documentclass[msmath,amssymb,aps,pra,twocolumn,epsfig,showpacs,bibliography,lengthcheck,superscriptaddress]{revtex4-2}

\usepackage{graphicx}
\usepackage{dcolumn}
\usepackage{bm}
\usepackage{hyperref}
\usepackage{epstopdf}
\usepackage{subcaption}
\usepackage{caption}
\usepackage{footmisc}
\usepackage{amsmath}
\usepackage{color}

\begin{document}
\title{A sensitive and stable atomic vector magnetometer for weak field detections using double orthogonal multipass cavities}%
\author{S.-Q. Liu}
\author{Q.-Q. Yu}
\author{H. Zhou}
\affiliation{Department of Precision Machinery and Precision Instrumentation, Key Laboratory of Precision Scientific Instrumentation of Anhui Higher Education Institutes, University of Science and Technology of China, Hefei 230027, China}

\author{D. Sheng}
\email{dsheng@ustc.edu.cn}
\affiliation{Department of Precision Machinery and Precision Instrumentation, Key Laboratory of Precision Scientific Instrumentation of Anhui Higher Education Institutes, University of Science and Technology of China, Hefei 230027, China}

\begin{abstract}
This paper presents a compact low-temperature atomic vector magnetometer for weak field measurements, using an atomic cell containing two orthogonal multipass cavities. At the working temperature of 75 $^\circ$C, the magnetic field sensitivities at all three axes are better than 45 fT/Hz$^{1/2}$ at 10~Hz limited by photon noise, and 85 fT/Hz$^{1/2}$ at 0.1~Hz. This sensor also shows measurement stabilities better than 1.5~pT at three axes for an integration time of $10^4$ s, even with the laser frequency unlocked. The sensor response to a rotation is demonstrated, which is also developed to measure the effective gyromagnetic ratio of atoms in this sensor when the bias field is nulled.  This magnetometer makes an important step towards long-term stable measurements and calibrations of ultra-low fields.
\end{abstract}
\maketitle

\section{Introduction}
Highly sensitive magnetometers have wide applications in fundamental physics researches~\cite{sachdeva2019,GNOME2021}, geology studies~\cite{nabighian2005,dang2010,prouty2013,yang2022}, and bio-imaging~\cite{zhangrui2020,limes2020}. Since the sensor sensitivity is determined by the signal-to-noise ratio (SNR) and coherence time of the studied system, there are several ways to improve the sensor sensitivity. One is to improve the atomic coherence time using coated cells~\cite{seltzer2013}, and the other one is to increase SNR by attaching multipass cells~\cite{Li2011} to the system. For alkali metal atoms, a combination of both improvements can be achieved in the high temperature spin-exchange-relaxation-free (SERF) regime, where the conditions of high atomic densities and suppressed transverse depolarization rate are simultaneously realized~\cite{allred02}.

Besides the field sensitivity, the long-term measurement stability is another important parameter of the magnetometer in practical applications, especially when it comes to precise comparisons of measurement results at different times. Among the aforementioned approaches, the formal two methods can work in a relatively low temperature, which helps to maintain a more stable measurement environment. Due to buffer and quenching gases, the atomic transition line widths in the multipass-cell method are largely broadened~\cite{romalis1997}, and the dependence of measurement results on the laser beam parameters is weaker compared with the coated cell method.

In this paper, we demonstrate a highly sensitive and stable atomic vector magnetometer for weak field measurements, with the laser free running, using the multipass-cell method. Following this introduction, the working principle and setup of the sensor are introduced in Sec. II, the results of sensor parameter calibrations and operation tests are presented in Sec. III, sensor field measurement stabilities are discussed in Sec.~IV, and the paper is concluded in Sec. V.

\section{Working Principle and Experiment Setup}
The magnetometer presented in this paper operates with the single-beam zero-field level-crossing resonance scheme~\cite{sheng17}. The dynamics of the electron spin polarization $\mathbf{P}$ is described by the Bloch equation~\cite{shah09}:
\begin{equation}
\label{eq:bloch}
\frac{d\mathbf{P}}{dt}=\frac{1}{Q(P)}[\gamma_e\mathbf{P}\times\mathbf{B}-R_{op}(\mathbf{s}-\mathbf{P})-R_{rel}\mathbf{P}],
\end{equation}
where $\gamma_e$ is the electron gyromagnetic ratio, $\mathbf{B}$ is the external field, $Q(P)$ is the nuclear-spin slowing-down factor~\cite{appelt98}, $R_{op}$ is the optical pumping rate, $\mathbf{s}$ is the photon spin, and $R_{rel}$ is the atomic depolarization rate in the absence of the laser beam. When the external field is expressed as $\mathbf{B}=\hat{v}[B_{v,0}+B_{v,m}\cos(\omega t)]$, with $\hat{v}$ perpendicular to the pump beam direction, the electron polarization along the pump beam direction is modulated at harmonics of $\omega$, with its first harmonic component as~\cite{cohen70,shah09}
\begin{equation}
\label{eq:Pw}
P(\omega)=\frac{\gamma_eB_{v,0}R_{op}\sin(\omega t)}{R^2_{tot}+(\gamma_e B_{v,0})^2}J_0\left(\frac{\gamma_e B_{v,m}}{   Q(P)\omega}\right)J_1\left(\frac{\gamma_eB_{v,m}}{Q(P)\omega}\right).
\end{equation}
Here, $R_{tot}=R_{op}+R_{rel}$, and $J_{0}$ ($J_{1}$) is the zero-order (first-order) Bessel function of the first kind. The intensity $I$ of the transmitted laser beam is dependent on $P$ according to the equation~\cite{appelt98}:
\begin{equation}~\label{eq:I}
\frac{dI(l)}{dl}=-I(l)[1-\mathbf{s}\cdot\mathbf{P}(l)]n\sigma,
\end{equation}
where $l$ is the propagation length of the light inside the atomic cell, $n$ is the atomic density, and $\sigma$ is the photon-absorption cross section of atoms. Similar to the atomic polarization, the transmitted beam intensity also shows correlated modulations from $B_{v,m}$, and we can extract the bias field along the modulation field direction by demodulating the first harmonic signal from the transmitted beam~\cite{sheng17}.

\begin{figure}[htb]
\includegraphics[width=3in]{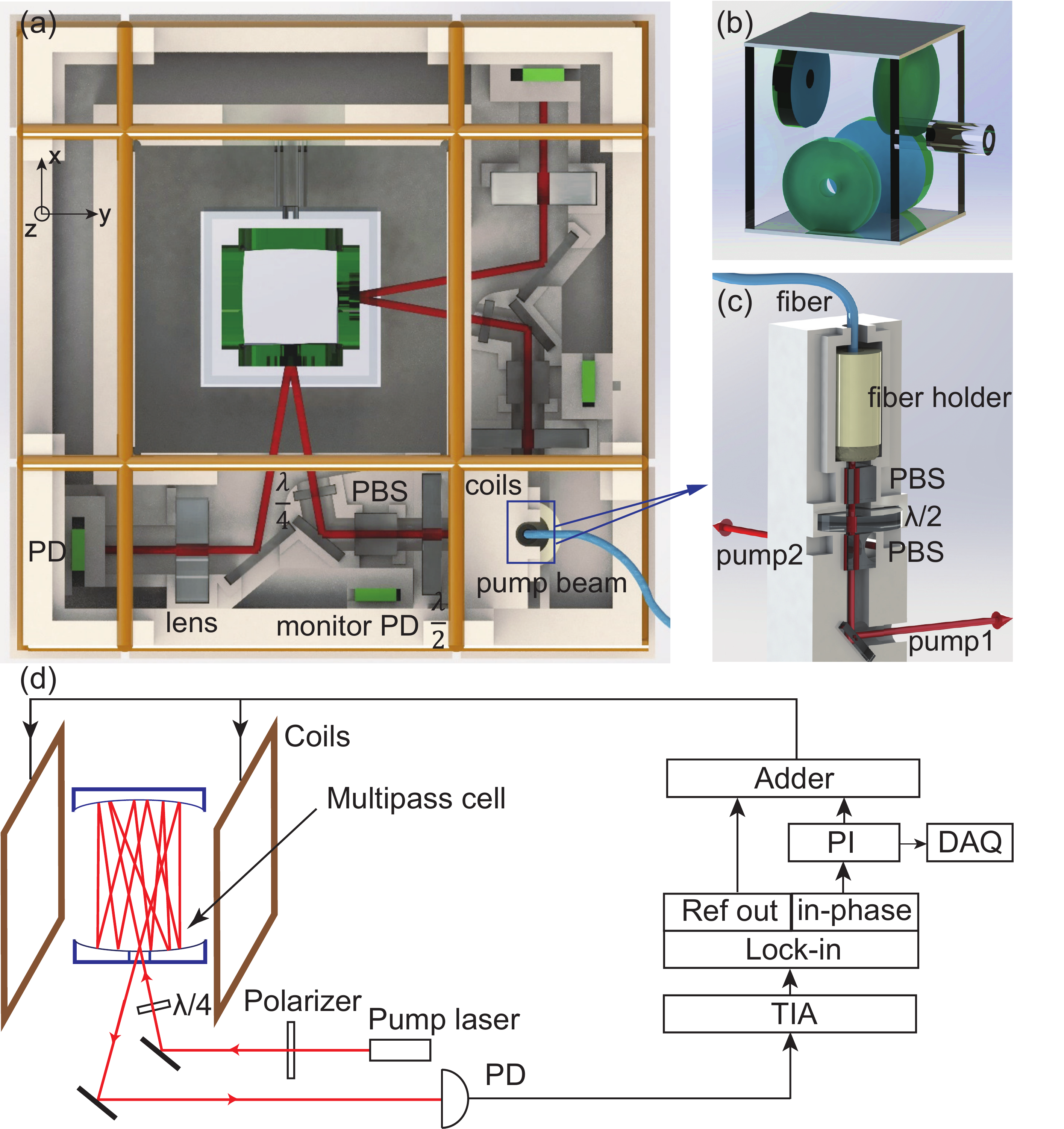}
\caption{\label{fig:setup}(a) Top view of the vector magnetometer sensor. (b) An illustration of the atomic cell used in the experiment, which contains two orthogonal multipass cavities. (c) Side view of the part that splits the beam from the fiber, where the upper PBS is used to purify the beam polarization, and the fiber hold is a 3D printed structure that contains a ceramic fiber ferrule and a collimation lens. (d) A schematic plot of the electrical signal processes.}
\end{figure}

For the single-beam zero-field level-crossing scheme, the sensor is insensitive to the magnetic field along the pump beam direction~\cite{boto2022,yan2022}. To eliminate this problem, we use a cell assisted by two Herriott cavities~\cite{silver2005} as illustrated in Fig.~\ref{fig:setup}(b). These two Herriott cavities are placed orthogonally to avoid the detection dead zones, and a vertical separation between the beam patterns inside the two cavities is kept to be larger than 3~mm, so that the effect of atomic diffusions from one cavity to the other during the coherence time can be neglected.  Such cavities are bonded on silicon wafers and attached to the glass cell using the anodic bonding technique~\cite{cai2020}. Each cavity consists of two identical cylindrical mirrors, while each mirror has a radius of curvature of 60~mm, and a diameter of 12.7~mm. The distance between the two mirrors is 11.5~mm, and the relative angle between their symmetrical axes is 52.2$^\circ$. There is a hole with a diameter of 2.5~mm in the center of the front mirror, and the optical beam enters and exits the cavity from the same hole with 21 reflections inside the cavity. The beam-direction for the magnetometer is defined by the averaged beam-orientation inside the cavity~\cite{yu2022}, which is along the connection of both cavity mirror centers, due to the symmetrical distributions of beams inside the cavity~\cite{silver2005}. In this way, the sensor beam-direction is insensitive to the input-beam orientation. The atomic cell, with an outer size of 21.7$\times$21.7$\times$25.5~mm$^3$, is filled with 600 torr N$_2$ gases and a droplet of $^{87}$Rb atoms. It normally operates at a temperature of 75 $^\circ$C by running high frequency ac currents through ceramic heaters, and the average heating power cost is 2.3~W, which is twice of the black-body radiation limit~\cite{sheng17}.

The optical platform of the magnetometer sensor, as shown in Fig.~\ref{fig:setup}(a), is produced by high-precision three-dimensionally (3D) printing technology. The sensor has a cubic shape with a side length 78.5~mm.  A beam, on resonance with Rb D1 transition line, is coupled into the sensor by a polarization maintaining fiber. This beam, with a diameter of 1~mm, is split into two beams by a polarizing beam splitter (PBS) for each Herriott cavity, as shown in Fig.~\ref{fig:setup}(c). Before entering the Herriott cavity, 10\% of each beam power is split to a photodiode detector (PD) for beam power monitoring, and the rest of the beam is circularly polarized by passing through a quarter-wave plate. The transmitted beam from each cavity is recorded by a PD. Modulation fields are generated by connecting the signals from lock-in amplifiers to three orthogonal pairs of Helmholtz coils. These coils are wrapped along the slots on the sensor outer surfaces, with identical side lengths of 74.6~mm. The modulation field in each direction has a similar amplitude around 150 nT, but each has a different modulation frequency, ranging from 1.8 kHz to 2.7 kHz. The specific modulation frequencies are chosen for the sensor to reach the best long-term stability, which will be discussed in the latter part of the paper.

All electrical signals pass through a printed circuit board, which is attached with the sensor. A shielded 12-core cable is used to transfer most of the electrical signals to the control electronics, except that the temperature signal is transmitted by a T-type thermal couple cable, and the heating current is carried by a separate shielded cable.  The current signal from the PD, recording the transmitted beam exiting from the atomic cell, is first converted to a voltage signal by a transimpedance amplifier (TIA), and then passed to the lock-in amplifier. The in-phase signal from the lock-in amplifier is sent to a feedback loop as shown in Fig.~\ref{fig:setup}(d), which is used to constantly null the bias field in each modulation field direction. In the closed-loop operations, the outputs from the proportional-integral (PI) controller are recorded as the magnetometer signals~\cite{sheng17}.

\section{Sensor Parameter Calibrations and Operation Tests}

\begin{figure}[htb]
\includegraphics[width=3in]{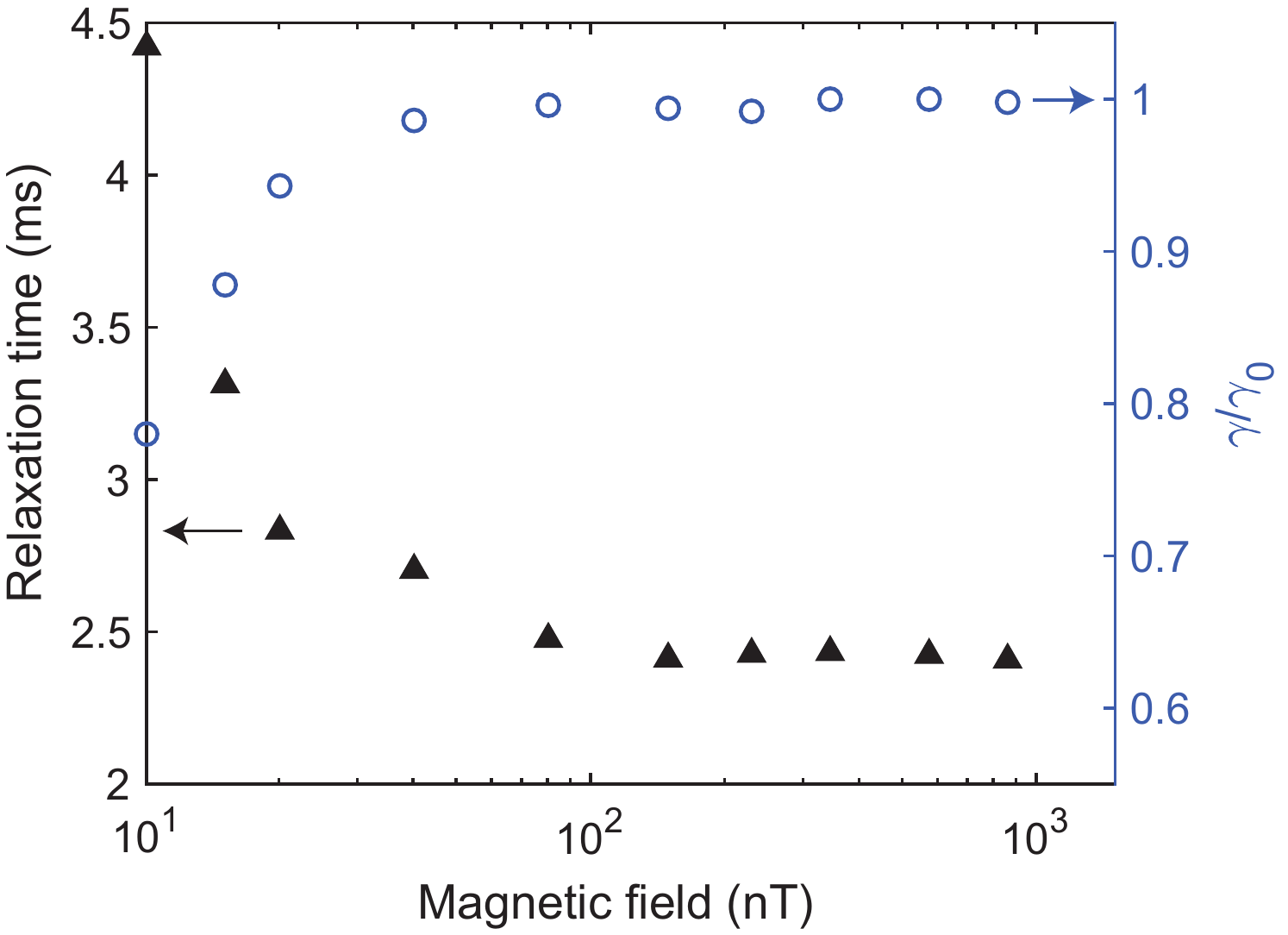}
\caption{\label{fig:relax} A plot of the atomic transverse relaxation time and atomic gyromagnetic ratio as a function of the bias field magnitude, with $\gamma_0$ as the atomic gyromagnetic ratio when the spin-exchange rate is negligible compared with the atomic Larmor precession frequency. All the data are taken with the sensor inside a 5-layer mu-metal shield, which is the case for all the data in the rest of the paper. }
\end{figure}

We first apply a pump-probe scheme~\cite{hao2019} to measure the parameters characterizing the atomic spin dynamics using a separate setup. Atoms are polarized by a resonant circularly polarized pump beam at zero bias field, and probed by an off-resonance linearly polarized beam in a transverse direction. Then the pump beam is turned off, and a bias field perpendicular to both the pump and probe beams is turned on. We record Faraday rotation of the probe beam polarization, from which the transverse depolarization time ($T_2$) and atomic gyromagnetic ratio  ($\gamma$) can be extracted.  As shown in Fig.~\ref{fig:relax}, $T_2$ is 2.4~ms when the bias field is larger than 80~nT. As the bias field is further reduced, the transverse relaxation is suppressed, together with the corresponding decrease of the atomic gyromagnetic ratio, which indicates that the system enters the SERF regime. Different from the high-temperature situation, the SERF mechanism in this case only helps to improve $T_2$ by a factor of two, due to the small spin-exchange rate in this study.

\begin{figure}[htb]
\includegraphics[width=3in]{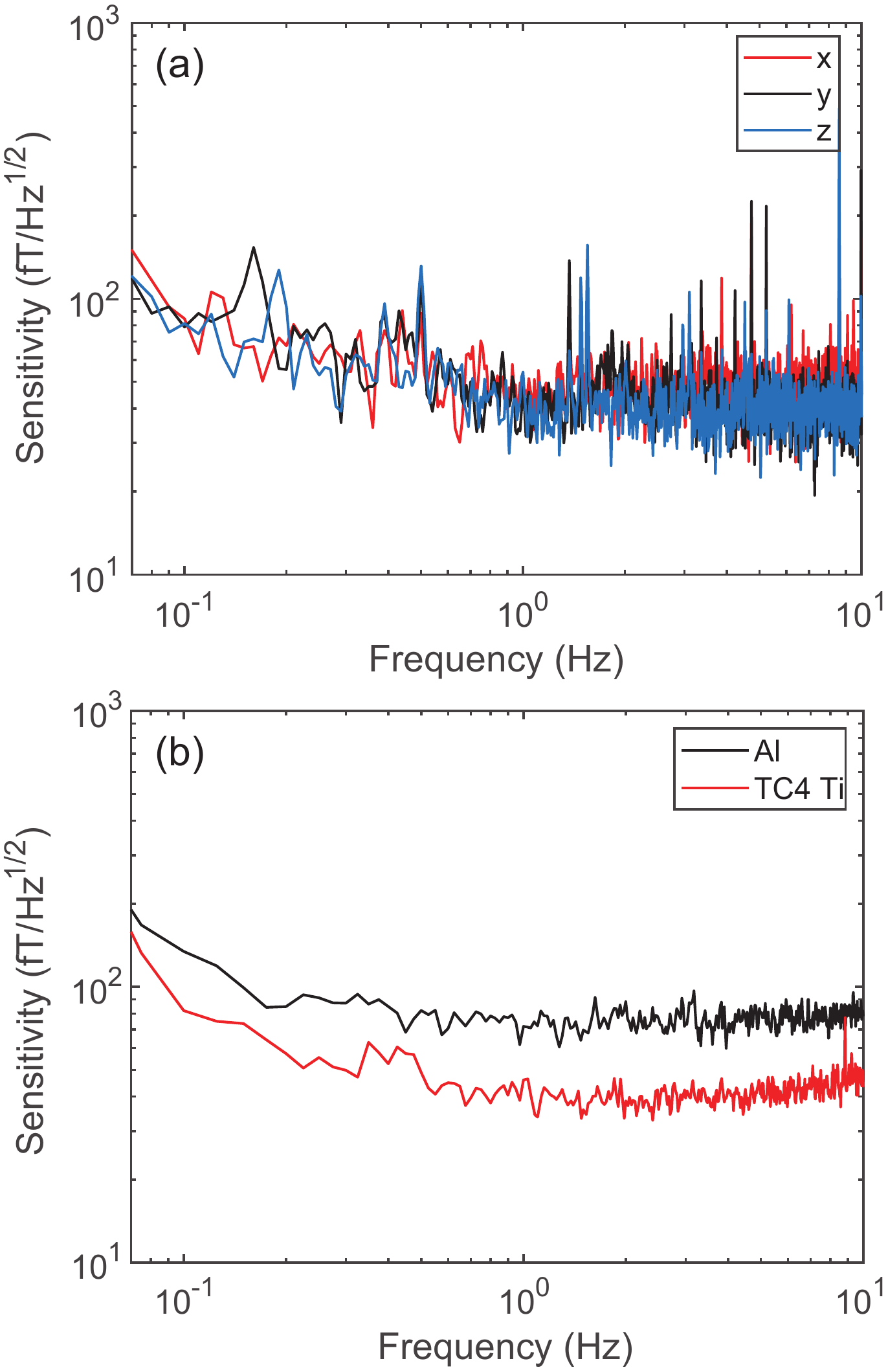}
\caption{\label{fig:sen} (a) A plot of the sensor magnetic field sensitivities simultaneously taken at all three axes. The close-loop bandwidth is set to be 30 Hz by the  time constant in the demodulation process, and it is kept the same in the rest of the paper. (b) Comparisons of magnetic field sensitivities along the $z$ axis when a square piece of TC4 titanium alloy or aluminum materials, with a thickness of 1.6~mm and a side length of 8~cm, is attached to the sensor in the $z$ direction.}
\end{figure}

With an injection beam power of 0.26 mW for each cavity, the transmitted beam power is around 0.1 mW, and the magnetometer open-loop response curve shows a full width at half maximum (FWHM) of 145 nT. This sensor has two independent closed-loop operation modes, since the signal from either multipass cavity can be used to feed back on the common sensitive magnetic field direction, $z$ axis in Fig.~\ref{fig:setup}(a). These two modes show similar magnetic field sensitivities.  Moveover, the sensor sensitivity reaches a plateau for the cell temperature above 75 $^\circ$C. In the rest of the paper, the magnetometer signals for $y$ and $z$ axes are extracted from the same probe beam, and the cell temperature is kept at 75$^\circ$C. As shown in Fig.~\ref{fig:sen}(a), magnetic field sensitivities along all three axes are below 45 fT/Hz$^{1/2}$ at 10~Hz, which are limited by the photon noise, and below 85 fT/Hz$^{1/2}$ at 0.1~Hz. Moreover, the cross-talks between different feedback channels are all measured to be less than 0.1\%.

From the considerations on the mechanical and electrical~\cite{seltzer2004} properties, it is preferred to pack the sensor head with metallic materials on the outside. In this case, it requires special attentions to the Johnson noise current in the metal~\cite{lee2008}. If a piece of metal disk is attached to the outside of the sensor in the $z$ axis in Fig.~\ref{fig:setup}(a), the resulted white noise felt by atoms in the $z$ direction is~\cite{lee2008}:
\begin{equation}~\label{eq:dB}
\delta B_z=\frac{1}{\sqrt{8\pi}}\frac{{\mu_0}\sqrt{k_BT\sigma t}}{d}\frac{1}{1+d^2/r^2}.
\end{equation}
Here, $\mu_0$ is the vacuum permeability, $k_B$ is the Boltzmann constant, $T$ is the environment temperature, $d$ is the distance between the center of the metal piece and atoms, $\sigma$, $t$ and $r$ are the electric conductivity, thickness and radius of the metal piece, respectively. For a square piece of commonly used metal, such as aluminum ($\sigma=3.8\times10^7$~Ohm$^{-1}\cdot$m$^{-1}$), with a thickness of 1.6~mm, a side length of 8~cm, and a distance of 28~mm from the center of the closer cavity among the two orthogonal ones, the resulted white noise for the magnetometer using the closer cavity is estimated to be 95~fT/Hz$^{1/2}$ from Eq.~\eqref{eq:dB}. The measurement result in Fig.~\ref{fig:sen}(b) corresponds to a field noise of 65 fT/Hz$^{1/2}$ from the metal piece after removing the photon noise contribution. Here, we overestimate the noise in the calculation due to the fact that Eq.~\eqref{eq:dB} only applies to the central region of the cavity, and its value decreases for the other parts of the cavity. Such a field noise from the metal piece is beyond the original sensitivity of the sensor in Fig.~\ref{fig:sen}(a), and it is needed to find metals that have smaller $\sigma$. One of possible replacements is TC4 titanium alloy (Ti-6Al-4V), which has an electrical conductivity of $\sigma=5.8\times10^5$~Ohm$^{-1}\cdot$m$^{-1}$. By changing the material used in the above measurement to the TC4 material, we can reduce the white noise from the packing material by a factor of 8, so that the original noise of the sensor dominates the field sensitivity in the frequency domain higher than 0.1 Hz. This is also confirmed by the result in Fig.~\ref{fig:sen}(b). However, we need to emphasize that, the field noise from TC4 material will overcome the original sensor noise in the frequency  region lower than 0.1~Hz. Therefore, the metallic package is not suitable for a long-term stability requirement.

\begin{figure}[htb]
\includegraphics[width=3in]{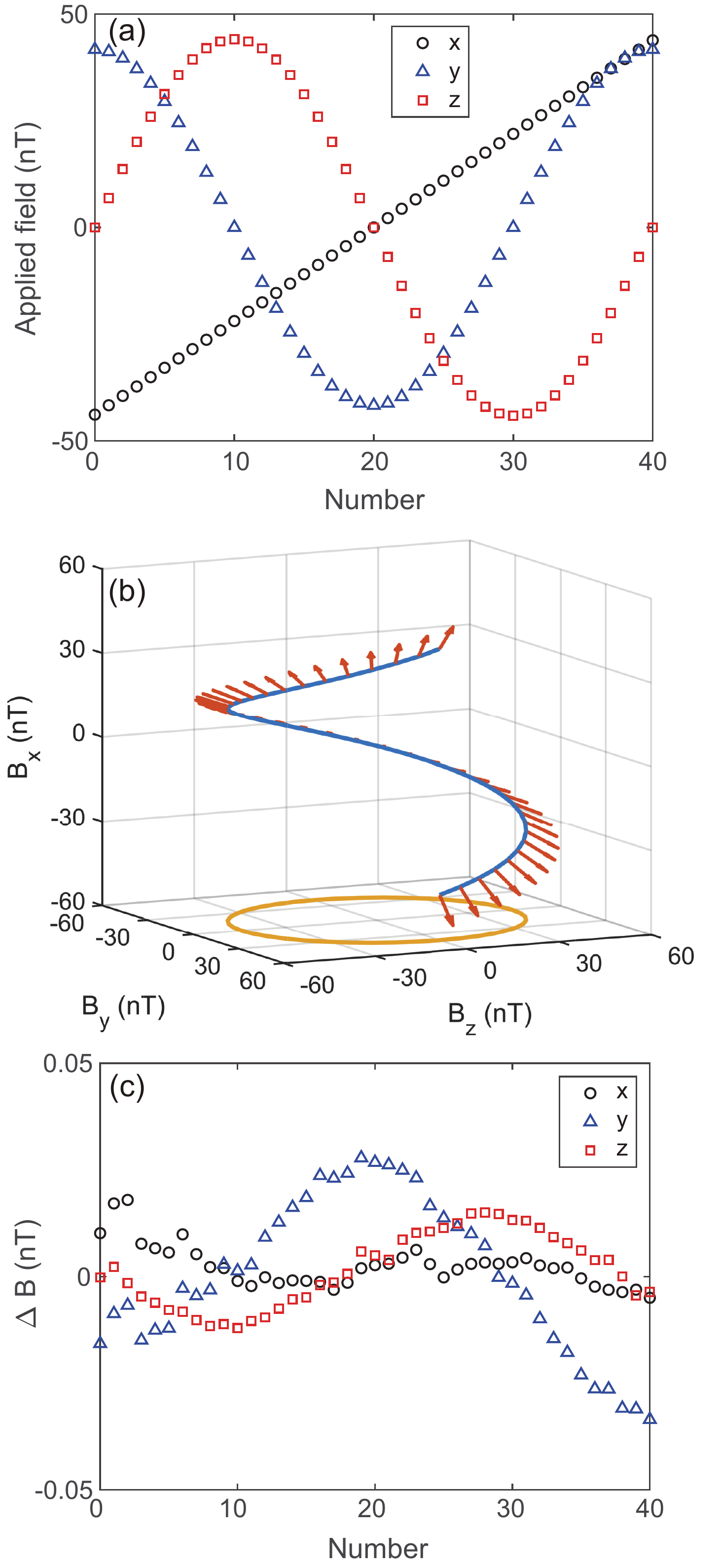}
\caption{\label{fig:vec} (a) 41 points of $\bm{B}_e$ used in the test of the vector magnetometer operation. (b) $\bm{B}_e$ visualized in the space of magnetic fields. The blue line is the curve formed by the 41 points, the red arrow is the direction of the field, and the yellow circle is the projection of the blue curve in the $B_y$-$B_z$ plane. (c) The discrepancy between the applied field and the sensor readout in three axes.}
\end{figure}

To demonstrate the operation of this three-axis vector magnetometer, we apply a series of external fields $\bm{B}_{e}$ with different orientations in the sensor coils as shown in Fig.~\ref{fig:setup}(d) when the sensor is working in the closed loop mode, and compare the applied field with the sensor readout. Following the scheme used in Ref.~\cite{zheng2020}, we choose 41 test points of $\bm{B}_e$ with the $i$th test field designed to be $B_{xi}=B_{dx}(\phi_i/\pi-1)$, $B_{yi}=B_{dy}\cos(\phi_i)$, and $B_{zi}=B_{dz}\sin(\phi_i)$, where $\phi_i=i/N\times2\pi$ and $N=40$. Figure~\ref{fig:vec}(a) shows the components of $\bm{B}_{e}$ with $B_{dx}$=43.87~nT, $B_{dy}$=41.72~nT, and $B_{dz}$=44.15~nT. The curve formed by these points in the space of magnetic fields is visualized in Fig.~\ref{fig:vec}(b). From the measurement, we find the discrepancy between the applied field and the sensor readout is within 0.1\% in all three axes.

\begin{figure}[htb]
\includegraphics[width=3in]{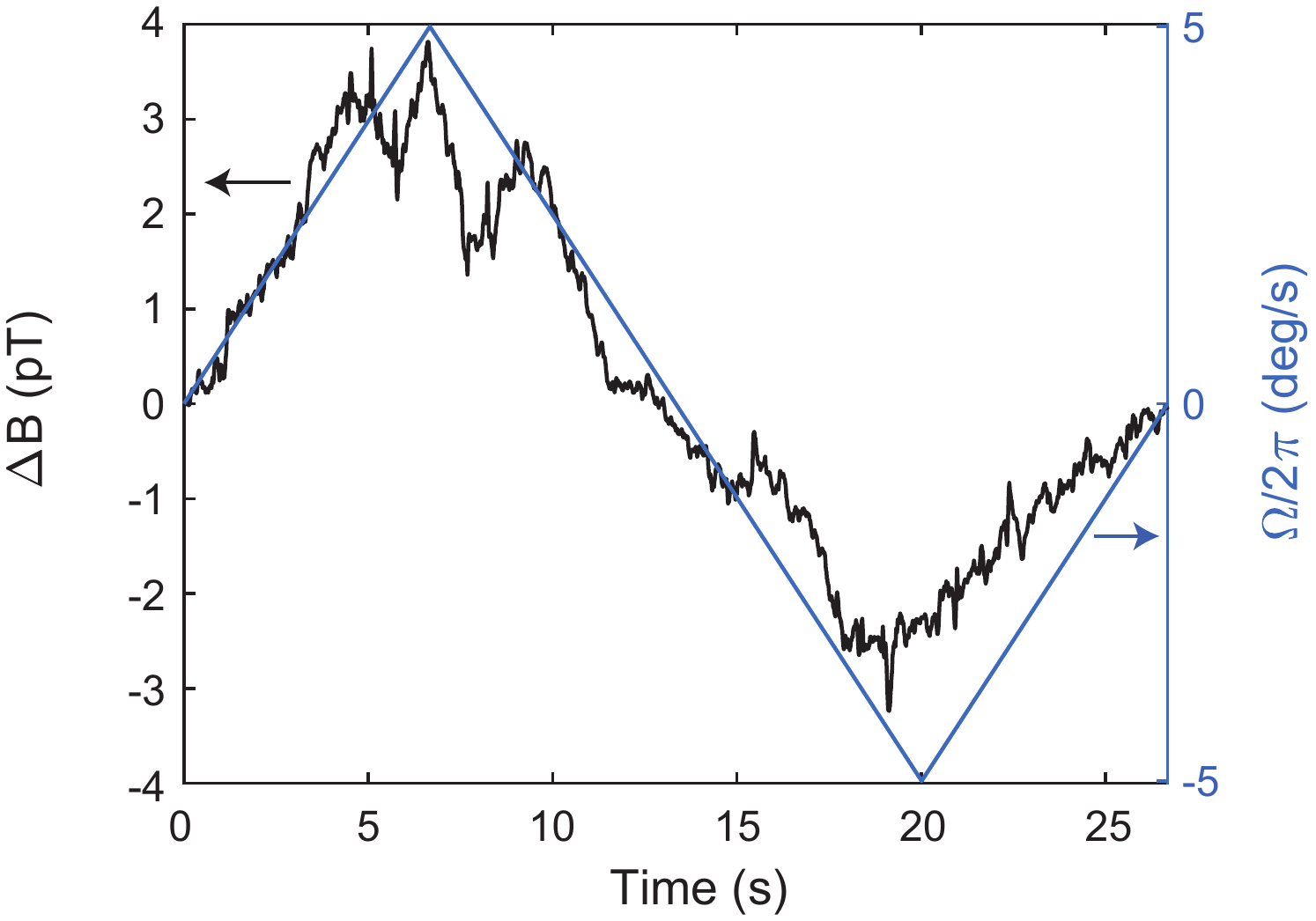}
\caption{\label{fig:acc} Time evolution of the rotation frequency of the test platform and the sensor response, where the rotation rate is under the same acceleration and deceleration rate. }
\end{figure}

As a gas spin sensor, this magnetometer is also sensitive to the external rotation, where a rotation with a frequency of $\Omega$ leads to an effective magnetic field of $\Delta B=\Omega/\gamma$ along the rotation axis. In this case, an external rotation can be used for in-situ calibration of $\gamma$. This provides an complementary method to the aforementioned ``free-induction-decay" method, which is not suitable to measure the effective atomic gyromagnetic ratio when the bias field is nulled in the closed loop mode. We place the test platform of the sensor on top of a motorized rotation table, and control the rotation rate with a constant acceleration and deceleration rate $|a|$ along the $y$ axis. As shown in Fig.~\ref{fig:acc}, the rotation frequency $\Omega$ evolves linearly in time and within the range of $\Omega_m/2\pi=5$~deg/s. The same figure also shows the recorded sensor signals along the $y$ axis, where the readout of the closed-loop magnetometer is synchronized with $\Omega$ as expected. By varying $|a|$ at several different values while keeping $\Omega_m$ the same, we extract a weighted result of $\gamma/\gamma_0=0.70\pm0.02$ at the null bias field.

\section{Sensor Measurement Stability}
\begin{figure}[htb]
\includegraphics[width=3in]{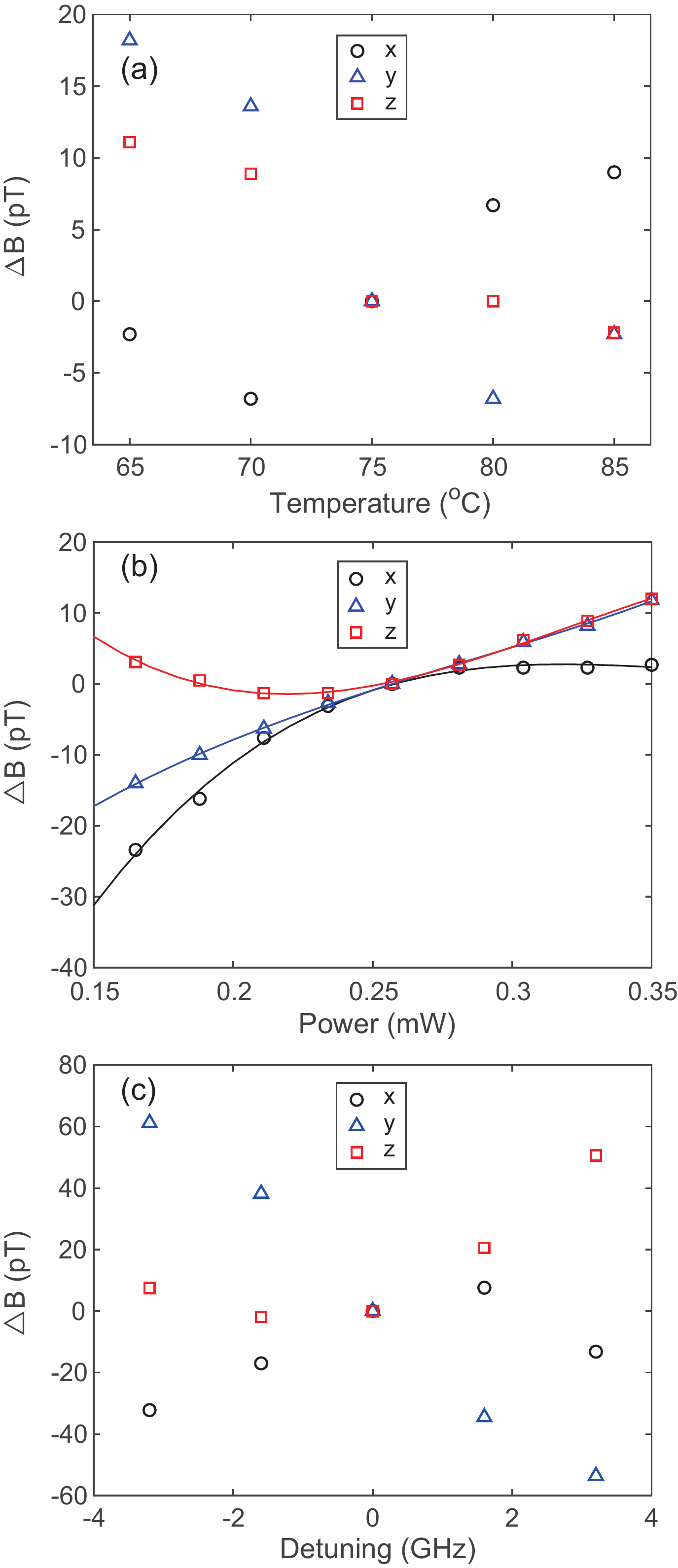}
\caption{\label{fig:stability} Plot (a), (b) and (c) show the dependence of the measurement results on the cell temperature, beam power before entering the multipass cavity, and beam frequency, respectively. The solid lines in plot (b) are fitting results using third order polynomial functions.}
\end{figure}

The magnetic field measurement results from this vector magnetometer are dependent on several experiment parameters. Besides the previously mentioned rotation effect, cell temperature is another important experiment parameter, which is directly correlated with the atomic density. Experiment results in Fig.~\ref{fig:stability}(a) show that, it requires a temperature stability better than 0.4 $^\circ$C for the sensor to reach a 1~pT measurement stability, which is relatively easy to achieve in practice using a temperature control system. We also noticed that, while the temperature dependence of the magnetometer results for $y$ and $z$ axes are similar, there are obvious discrepancies between them and the result in the $x$ axis. This is partly due to the fact that the signals for the $y$ and $z$ axes are extracted form the same probe beam, while the signal for $x$ axis is from a different probe beam.

Laser parameters, including the beam power and frequency, are dominant experiment factors for the measurement stability. Their influences on the measurement come from the combined effects due to changes in optical pumping rates, light shifts and other technical reasons.  As shown in Fig.~\ref{fig:stability}(b), the changes in the field measurement results can be more than 10 pT, when the fluctuation of the beam power is beyond 20\%. These beam-power dependence can be well fitted by third order polynomial functions. Using these fitting results, together with the beam power monitor signals from the sensor, we can correct, and therefore, eliminate the beam-power effect on the magnetometer measurement results even without locking the beam power. For the distributed-Bragg-reflector (DBR) diode laser (Photodigm PH795DBR diode laser, which has a TO-8 package and a maximum output power of 80 mW) used in this paper, the laser frequency typically fluctuates by 200~MHz in a day, with ambient temperature changing by several degrees.

\begin{figure}[htbp]
\includegraphics[width=3in]{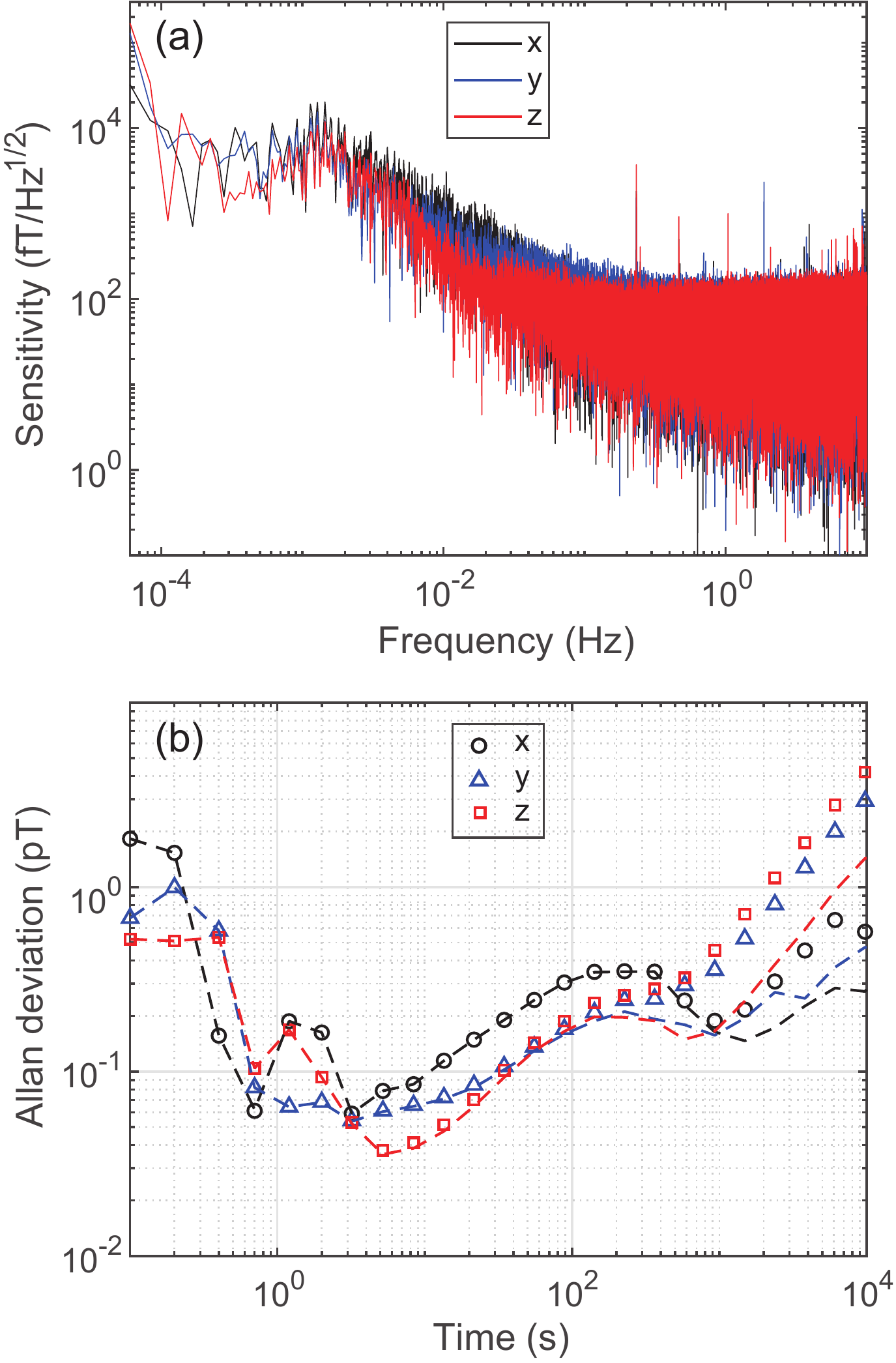}
\caption{\label{fig:allan} Plot (a) is the spectra density plot for a 10-hour-data long magnetometer data. The bump around 1~mHz is caused by the instability of the cell temperature which is controlled based on a T-type thermal couple. Plot (b) is the Allan deviation results of the measurement results in plot~(a) with (dashed lines) and without (symbols) corrections of the beam power effect. }
\end{figure}

To measure the stability of this sensor, we run the magnetometer in the closed-loop mode overnight without any control on the laser beam parameters. Data points in Fig.~\ref{fig:allan}(a) and (b) show the analysis of a 10-hour-long magnetometer data in both frequency and time domain. With an integration time around 3 hours, the stability of the measured field in $x$ axis is still better than 1 pT, but the results in  $y$ and $z$ axes reach 3 and 4 pT, respectively. These results from the raw data can be improved by more than a factor of two once the beam-power effect is corrected. The resulted data sets, which are shown as dashed lines in the same plot, demonstrate that the measurement stabilities in all three axes are better than 0.2~pT for an integration time of 10$^3$~s, which surpass that of the vector magnetometer at zero field using a coated cell~\cite{bison2018}. With an integration time of $10^4$~s, the measurement stabilities are better than 1.5~pT for all three axes, among which the best one is better than 0.3 pT. This performance is comparable with the state-of-art scalar magnetometers used in the search for neutron electric dipole moment~\cite{abel2020}. For the time scale presented in Fig.~\ref{fig:allan}(b), the laser frequency drift is not yet an important source of the measurement instability.

\begin{figure}[htbp]
\includegraphics[width=3in]{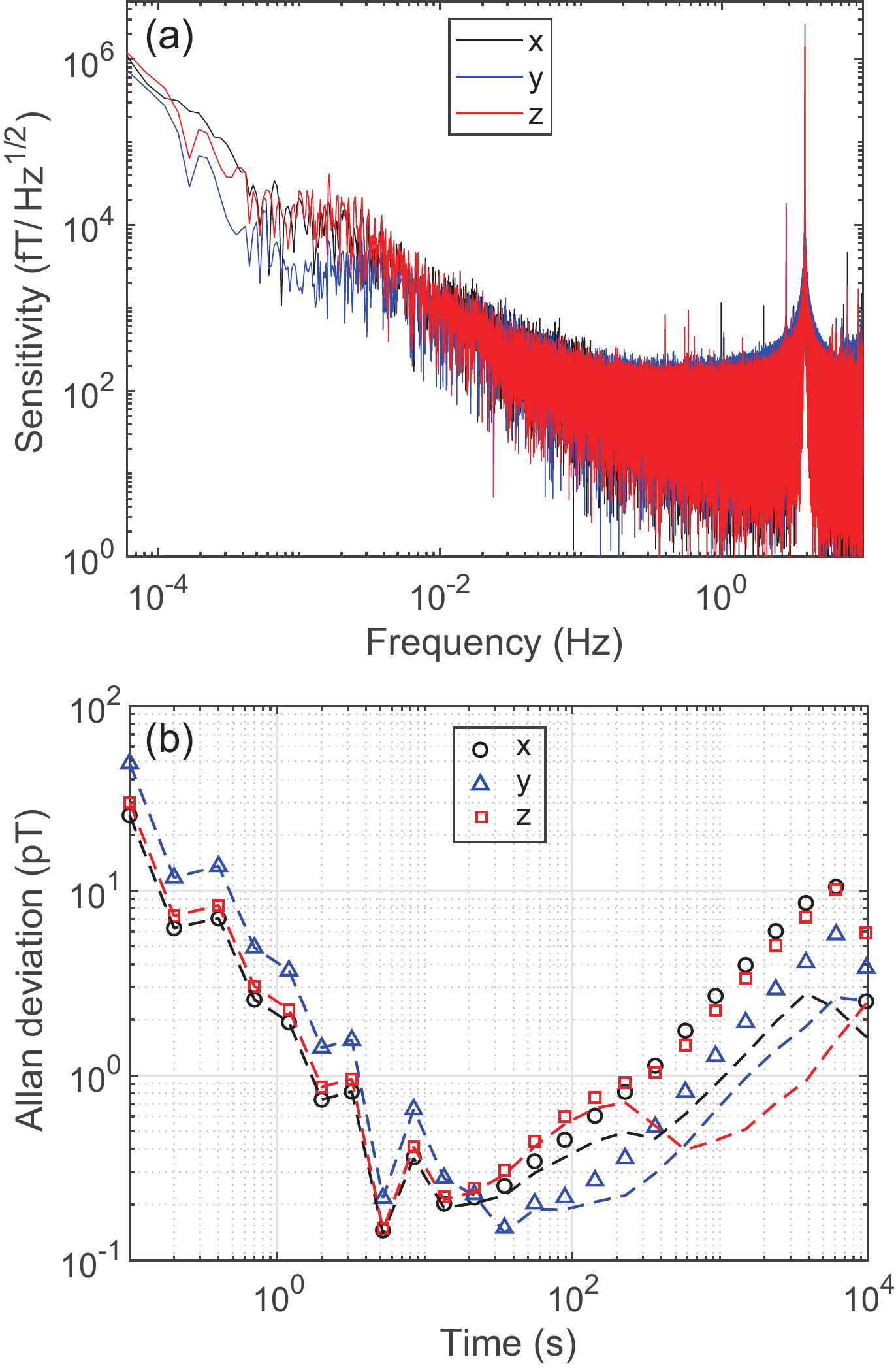}
\caption{\label{fig:allan2} Plots are same as the ones in Fig.~\ref{fig:allan}, except that the beating signals of the modulation fields appear in the frequency domain below 10~Hz.}
\end{figure}

It needs to be emphasized that one additional important condition for the measurement results presented in Fig.~\ref{fig:allan}(b) is that, the beating frequencies of the modulating fields and several of their main harmonics are beyond the detection bandwidth of the sensor. Otherwise, as shown in Fig.~\ref{fig:allan2}, the long-term stability of the sensor will be disturbed. There are also several other schemes to realize similar closed-loop operations of the sensor. In these alternatives, the modulation fields in multiple directions share the same frequency, and are different only in phases~\cite{boto2022}. By comparing different schemes, we find from the measurements that current scheme, where the modulation fields in three directions are independent, suffers most significantly from the frequency-beating problem. However, once care is taken as in the aforementioned measurements, we also find that it is this independent-modulation-field scheme used in this paper that helps the sensor to reach the best long-term measurement stabilities in all three axes.

\section{Conclusion}
In conclusion, we have demonstrated a compact atomic vector magnetometer assisted by double orthogonal multipass cavities. It works in an environment with low temperature and buffer gases, and maintains a weak field measurement stability better than 1.5 pT at a integration time of 10$^4$~s even without laser locking. This device is also promising to be extended to operate in the earth field range~\cite{sheng2013,cai2020} with suitable modifications.

Several aspects of this sensor can be improved. One is to further reduce the sensor size and power cost, which can be achieved by decreasing the cavity length, and in turn, the cell size. The other part is on the measurement stability. It is reasonable to expect a measurement stability around 1 pT over the course of days or even longer, and such a measurement stability is a necessary condition for the ultra-low field calibration. In this time scale, it needs to implement more controls on the laser parameters. We have worked out a new scheme for beam power locking with a low cost of power and space, and a robust way to avoid the modulation-frequency-beating problem~\cite{zhouprep}. For the beam frequency locking, there are already some well-established methods~\cite{demtroder2008,lee2011}. However, for the purpose of ultimate miniaturation, it may be better to implement a pulse scheme with the beams already in the sensor, considering the fact that we only need to feed back on the beam frequency around every several hours.


\section*{Acknowledgements}
This work was partially carried out at the University of Science and Technology of China (USTC) Center for Micro and Nanoscale Research and Fabrication. This work was supported by the National Natural Science Foundation of China (NSFC, Grant No. 11974329), and by the Scientific Instrument and Equipment Development Projects, Chinese Academy  of Sciences (CAS, Grant No. YJKYYQ20200043).

\end{document}